\documentclass[aps, prd, twocolumn, superscriptaddress, nofootinbib]{revtex4}
\usepackage{graphicx}
\usepackage{dcolumn}
\usepackage{amssymb,amsmath,amsthm,mathrsfs}
\usepackage{epstopdf}

\begin{document}

\newcommand{\Eq}[1]{Eq. \ref{eqn:#1}}
\newcommand{\Fig}[1]{Fig. \ref{fig:#1}}
\newcommand{\Sec}[1]{Sec. \ref{sec:#1}}

\newcommand{\PHI}{\phi}
\newcommand{\vect}[1]{\mathbf{#1}}
\newcommand{\Del}{\nabla}
\newcommand{\unit}[1]{\mathrm{#1}}
\newcommand{\x}{\vect{x}}
\newcommand{\ScS}{\scriptstyle}
\newcommand{\ScScS}{\scriptscriptstyle}
\newcommand{\xplus}[1]{\vect{x}\!\ScScS{+}\!\ScS\vect{#1}}
\newcommand{\xminus}[1]{\vect{x}\!\ScScS{-}\!\ScS\vect{#1}}
\newcommand{\diff}{\mathrm{d}}

\newcommand{\be}{\begin{equation}}
\newcommand{\ee}{\end{equation}}
\newcommand{\bea}{\begin{eqnarray}}
\newcommand{\eea}{\end{eqnarray}}
\newcommand{\vu}{{\mathbf u}}
\newcommand{\ve}{{\mathbf e}}
\newcommand{\vk}{{\mathbf k}}
\newcommand{\vx}{{\mathbf x}}
\newcommand{\vy}{{\mathbf y}}

\newcommand{\uden}{\underset{\widetilde{}}}
\newcommand{\den}{\overset{\widetilde{}}}
\newcommand{\denep}{\underset{\widetilde{}}{\epsilon}}

\newcommand{\nn}{\nonumber \\}
\newcommand{\dd}{\diff}
\newcommand{\fr}{\frac}
\newcommand{\del}{\partial}
\newcommand{\eps}{\epsilon}
\newcommand\CS{\mathcal{C}}

\def\be{\begin{equation}}
\def\ee{\end{equation}}
\def\ben{\begin{equation*}}
\def\een{\end{equation*}}
\def\bea{\begin{eqnarray}}
\def\eea{\end{eqnarray}}
\def\bal{\begin{align}}
\def\eal{\end{align}}


\title{Chirality of tensor perturbations for complex values of the Immirzi parameter}

\newcommand{\addressImperial}{Theoretical Physics, Blackett Laboratory, Imperial College, London, SW7 2BZ, United Kingdom}

\author{Laura Bethke}
\affiliation{\addressImperial}

\author{Jo\~{a}o Magueijo}
\affiliation{\addressImperial}


\date{\today}

\begin{abstract}
In this paper we generalise previous work on tensor perturbations in a de Sitter background in terms of Ashtekar variables to cover all complex values of 
the Immirzi parameter $\gamma$ (previous work was restricted to 
imaginary  $\gamma$). 
Particular attention is paid to the case of real $\gamma$. Following the same approach as in the imaginary case, we can obtain physical graviton states by invoking reality and torsion free conditions. The Hamiltonian in terms of graviton states has the same form whether $\gamma$ has a real part or not; however changes occur for the vacuum energy and fluctuations. Specifically, we observe a $\gamma$ dependent chiral asymmetry in the vacuum fluctuations only if $\gamma$ has an imaginary part. Ordering prescriptions also change this asymmetry. We thus present a measurable result for CMB polarization experiments that could shed light on the workings of quantum gravity.
\end{abstract}

 \pacs{To be done}

\maketitle


\section{Introduction}
Although loop quantum gravity~\cite{pulbook,rovbook,thbook}  
is endowed with a rigorous mathematical structure,
it is still difficult to obtain GR as a low-energy limit from it and make contact with experiments. 
However, progress has recently been made on the computation of the graviton 
propagator~\cite{recentgravitons,rovrecent}, 
and in a previous publication ~\cite{prl} we have identified graviton states within the Hamiltonian framework
for a self-dual (or anti-self-dual) connection (for which the Immirzi parameter is $\gamma=\pm i$) . The detailed calculation for general imaginary values of $\gamma$ was provided in~\cite{paper}. To identify the graviton states that correspond to the dynamical, fluctuating part of space-time we compared our approach to cosmological perturbation theory. After taking several subtleties into account (for more details see~\cite{paper}) the Ashtekar Hamiltonian indeed reduces on-shell to the standard tensor perturbation Hamiltonian~\cite{muk}. But novelties come about. We found that only half of the graviton states are physical, retaining only the standard two polarisations for gravitons after reality conditions are imposed. For the physical states we discovered a $\gamma$-dependent chirality in the vacuum energy as well as the 2-point function.

In this paper, these results will be generalised to complex $\gamma$ in a Lorentzian theory. This is a non-trivial algebraic exercise with significant modifications in the results for the intermediate steps, but the final result is remarkably simple. 
For details on how to derive the second order Hamiltonian for gravitons Reference~\cite{paper} should be consulted; here we just summarize the framework and highlight the changes that occur for general $\gamma$. These are most notably the reality conditions and commutation relations between the canonical variables. 
It turns out that, in spite of these modifications, the final result is very simple: The vacuum chirality derived in~\cite{prl,paper} is only present if $\gamma$ has an imaginary part; for real $\gamma$ the two graviton polarisations are symmetric. 

The plan of the paper is as follows. 
In Section II we introduce the perturbed metric and connection variables and their classical solution. Section III explains the reality conditions and commutation relations for general $\gamma$. We present a representation of the Hamiltonian in terms of graviton states in section IV. In section V we explain how a complex $\gamma$ leads to a chirality in the vacuum fluctuations, but only provided that $\gamma$ has an imaginary part. The special case of real $\gamma$ will be investigated in section VI. We finish with a concluding section summarising our results.

\section{Notation and classical solution}
In this Section we lay down the notation, referring the reader to
previous publications~\cite{prl,paper} for details. 
We consider tensor fluctuations around de Sitter space-time 
described in the flat slicing,
$ds^2=a^2[-d\eta^2 +(\delta_{ab}+h_{ab})dx^adx^b] $, where
$h_{ab}$ is a symmetric TT tensor, $ a=- 1/H\eta$,
$H^2=\Lambda/3$ and $\eta<0$. 
Using the convention $\Gamma^i=-\frac{1}{2}\epsilon^{ijk}\Gamma^{jk}$
(where $\Gamma^{ab}$ is the spin connection), the 
Ashtekar-Immirzi-Barbero connection is given by
$A^i=\Gamma^i+\gamma \Gamma^{0i}$, with $\gamma$ the Immirzi parameter. 
Making use of the Cartan equations
for the zeroth order solution, the canonical variables can be expressed as: 
\bea\label{pertA} A^i_a&=&\gamma Ha \delta^i_a + \frac{a^i_a}{a}\\
E^a_i&=&a^2\delta^a_i - a\delta e^a_i \label{pertE}\; ,\eea where
$E^a_i$ is the densitized inverse triad, canonically conjugate to
$A^i_a$. As in~\cite{prl,paper} we define $\delta e^i_a$ via the triad,
$e^i_a=a\delta^i_a+\delta e^i_a$; we then raise and lower indices
in all tensors with the Kronecker-$\delta$, possibly mixing group
and spatial indices. This simplifies the notation
and is unambiguous if it's understood that $\delta e$ is
originally the perturbation in the triad.
It turns out that $\delta e_{ij}$ is proportional
to the ``$v$'' variable used by cosmologists~\cite{muk,lid}.

The canonical variables have symplectic structure
\be\label{PBnonpert}\{A^i_a(\vx),E^b_j(\vy)\}=\gamma
l_P^2\delta^b_a\delta^i_j\delta(\vx-\vy)\;  \ee which implies~\cite{paper}
\be\label{PBpert}\{a^i_a(\vx),\delta e^b_j(\vy)\}=-\gamma
l_P^2\delta^b_a\delta^i_j\delta(\vx-\vy)\; \; . \ee 
To make contact with cosmological perturbation theory 
and standard perturbative quantum field theory we use 
mode expansions (see~\cite{paper} for a full explanation):
\bea \delta
e_{ij}&=&\int \frac{d^3 k}{(2\pi)^{\frac{3}{2}}} \sum_{r}
\epsilon^r_{ij}({\mathbf k}) {\tilde e}_{r+}(\vk,\eta)e^{i\vk\cdot \vx}
\nonumber\\
&&
+\epsilon^{r\star}_{ij}({\mathbf k}) {\tilde e}^{\dagger}_{r-}(\vk,\eta)
e^{-i\vk\cdot \vx}\nonumber\\
a_{ij}&=& \int \frac{d^3 k}{(2\pi)^{\frac{3}{2}}} \sum_{r}
\epsilon^r_{ij}({\mathbf k}) {\tilde a}_{r+}(\vk,\eta)e^{i\vk\cdot \vx}
\nonumber\\
&& +\epsilon^{r\star}_{ij}({\mathbf k}) {\tilde a}^{\dagger}_{r-}(\vk,\eta)
e^{-i\vk\cdot \vx} \label{fourrier}\eea where ${\tilde e}_{rp}(\vk,\eta)=e_{rp}(\vk) \Psi_e(k, \eta)$ 
and ${\tilde a}_{rp}(\vk,\eta)=a_{rp}(\vk) \Psi^{rp}_a(k, \eta)$,
and $\epsilon^r_{ij}$ are polarization tensors. Amplitudes 
${\tilde a}_{rp}(\vk)$ and ${\tilde e}_{rp}(\vk)$ 
have two indices (contrasting  with previous literature, 
e.g.~\cite{gravitons,leelaur}):
$r=\pm 1$ for right and left helicities, and $p$ for graviton ($p=1 $) 
and anti-graviton ($p=-1$) modes.  
The $a_{rp}$ and $e_{rp}$  can be chosen so as {\it not} 
to carry any time-dependence, and for simplicity we will assume that they 
are equal. After imposing on-shell conditions  we'll find that 
functions $\Psi_a(k,\eta)$ must then carry an $r$ and $p$ 
dependence.

The classical solution  in terms of these variables can be 
read off from cosmological perturbation theory~\cite{paper}.
Since $\Psi_e$ is proportional to the ``$v$'' variable used in 
Cosmology~\cite{muk,lid}, it must satisfy the well-known equation
$\Psi_e''+{\left(k^2-\frac{2}{\eta^2}\right)}\Psi_e=0$ where
$'$ denotes derivative with respect to conformal time. This has
solution: \be\label{psie} \Psi_e=\frac{e^{-ik \eta}}{2 \sqrt  { k} }
{\left( 1-\frac{i}{k\eta} \right)}\; ,\ee where the normalization
ensures that the amplitudes $e_{rp}$  become
annihilation operators upon quantization. 
The connection can then be inferred from 
Cartan's torsion-free condition $ T^I=d e^I + \Gamma^I_J\wedge e^J=0$.
To first order, this is solved by \bea
\delta\Gamma^{0}_{\;i}&=& \frac {1}{a}{\delta e'_{ij}} \, dx^j \\
\delta \Gamma_{ki}&=&-\frac{2}{a}\partial_{[k}\delta e_{i ] j}\,
dx^j\label{gammaij}\; . \eea These imply $\delta \Gamma^{i}=\frac{1}{a}\epsilon
^{ijk}\partial_{j}\delta e_{kl}\, dx^l$, so that
\be\label{arealsp} a_{ij}=\epsilon_{ikl}\partial_k\delta
e_{lj}+\gamma\delta{e}'_{ij} \; .\ee Up until this point 
the calculation
is valid for all complex $\gamma$. The first novelty in this paper
appears upon inserting decomposition (\ref{fourrier}) into (\ref{arealsp}),
to determine torsion-free conditions in Fourier space.
Using relation 
$\epsilon_{nij}\epsilon^{r}_{il}k_j= i r k \epsilon ^{r}_{nl} $ we
obtain: \bea\label{psiapm} \Psi_a^{r+}&=&\gamma \Psi'_e + r
k\Psi_e\\ 
\Psi_a^{r-}&=&\gamma^\ast \Psi'_e + r
k\Psi_e\; , \eea
and clearly $\gamma^\star =-\gamma$, used in~\cite{prl}, is only
valid if $\gamma$ is imaginary. 
By writing a generally complex $\gamma$ as \be \label{gamma} \gamma=
\gamma_R+i\gamma_I \ee
we find that inside the horizon ($|k\eta|\gg1$)
\be \label{Psia} \Psi_a^{rp}=\Psi_ek\left(r-i\gamma_R+p\gamma_I \right)\; , \ee
generalizing the expression derived in~\cite{paper}. 
We note that the $p$ dependence of these functions only occurs if $\gamma$ has 
an imaginary part. For a real $\gamma$, $\Psi_a$ is the same for both
gravitons and anti-gravitons, as expected (a real connection would
be expanded in terms of a single particle ${\tilde a}_r$, 
so an index $p$ would be unnecessary; see Section~\ref{realgamma}
for a longer discussion). 
This is a first hint that the chirality
found in~\cite{prl,paper} is specific to non-real $\gamma$.

\section{Reality conditions and Commutation relations}

To be able to relate graviton and anti-graviton states 
(and their respective Hermitian conjugates), we need to impose 
reality conditions. As in~\cite{paper}, this will be done via the choice of inner product,
rather than as operator conditions. Nonetheless it
is important to see what these conditions look like in terms of operators
(or as classical identities). As the metric 
is real ($\delta e_{ij}=\delta e^\dagger_{ij}$), we have
\be \label{realg}e_{r+}(\vk)=e_{r-}(\vk) \; . \ee
The definition of the connection implies
\bea \label{reality}  \Re A^i&=&\Gamma^i+\gamma_R \Gamma^{0i} \\
\Im A^i&=&\gamma_I\Gamma^{0i}\; . \eea
Compared to the corresponding expressions for imaginary 
$\gamma$ (see~\cite{paper}), we note that the real part of 
the connection now has a contribution from $\Gamma^{0i}$, i.e. the 
extrinsic curvature. The reality conditions for the connection
should embody the non-dynamical torsion-free conditions, 
i.e. those not involving the extrinsic curvature, which in the
Hamiltonian formalism becomes the time derivative of the metric.  
The full torsion-free conditions representing (\ref{arealsp}) are now:
\bea a_{ij}+{\overline a}_{ij}&=&2a \left(\delta \Gamma_{ij}+\gamma_R \delta \Gamma^0_{ij}\right) \nonumber \\ &=& 2 \epsilon
_{ikl}\partial_k \delta e_{lj}+2\gamma_R \delta e'_{ij}\\
a_{ij}-{\overline a}_{ij}&=&2a i\gamma_I \delta \Gamma^0_{ij} =2i\gamma_I \delta e'_{ij}\; , \eea
or, in terms of Fourier components:
\bea\label{modereal} {\tilde a}_{r+} (\vk,\eta)+ {\tilde a}_{r-}(\vk,\eta) &=& 2 r k 
{\tilde e}_{r+}(\vk,\eta)+2 \gamma_R{\tilde e}'_{r+}(\vk,\eta)\, \, \,\, \, \, \, \, \, \,\,\,\\
\label{modereal2}
{\tilde a}_{r+} (\vk,\eta)- {\tilde a}_{r-}(\vk,\eta) &=& 2 i \gamma_I {\tilde e}'_{r+}(\vk,\eta)\; .
\eea 
Combining (\ref{modereal}) 
and (\ref{modereal2}) so as to eliminate the time derivative in the metric
leads to the condition:
\be\label{totreal} i\gamma^\ast{\tilde a}_{r+} (\vk,\eta)- i\gamma{\tilde a}_{r-}(\vk,\eta) = 
2 r k \gamma_I {\tilde e}_{r+}(\vk,\eta)\; .\ee
Its Hermitian conjugate is:
\be\label{totrealdagger} -i\gamma {\tilde a}^\dagger_{r+} (\vk,\eta)+ i\gamma^
\ast{\tilde a}^\dagger_{r-}(\vk,\eta) = 
2 r k \gamma_I {\tilde e}^\dagger_{r-}(\vk,\eta)\; ,\ee
which also invokes (\ref{realg}). 
These expressions represent the reality conditions that should
be imposed quantum mechanically by the choice of inner product. 
They are very different from their counterparts for a purely
imaginary $\gamma$ and represent novelty number two in our calculation.
For each $r$ and $\vk$ there are two independent conditions upon the
four operators $a_{rp}(\vk)$ and $e_{rp}(\vk)$. In addition to them there is
an independent dynamical torsion-free condition.
On shell, i.e. using (\ref{Psia}) and invoking (\ref{realg}), the connection
can be written in terms of the metric according to the weak identity:
\bea\label{modesonshell-} {\tilde a}_{r-}(\vk,\eta)&\approx&
rk {\tilde e}_r +\gamma^\ast {\tilde e}_r'\rightarrow
{\tilde e}_r(r-i\gamma^\ast)k \nonumber \\
{\tilde a}_{r+}(\vk,\eta)&\approx&
rk {\tilde e}_r +\gamma {\tilde e}_r'\rightarrow
{\tilde e}_r(r-i\gamma)k\; ,\label{modesonshell+} \eea
where the latter expression is valid in the limit $k|\eta|\gg 1$. These will
be useful in deriving the graviton operators for this theory. They
render one of the graviton modes unphysical, fully
relating metric and connection.

Before we can set up a quantum theory in terms of graviton operators we need
to define the commutation relations in terms of modes. These are obtained, 
as usual,
from the Poisson brackets (\ref{PBnonpert}) and (\ref{PBpert}),
leading to:
\be\label{unfixedcrs} \left[A^i_a(\vx),E^b_j(\vy)\right] = i\gamma
l_P^2\delta^b_a\delta^i_j\delta(\vx-\vy)\;  \ee
and \be\label{unfixedcrs1} \left[a^i_a(\vx),\delta
e^b_j(\vy)\right] = -i\gamma
l_P^2\delta^b_a\delta^i_j\delta(\vx-\vy)\; . \ee
The commutators for the mode expansions can be derived as in \cite{paper} 
and are:
\be\label{fixedcrs} [{\tilde a}_{rp}(\vk),{\tilde
e}_{sq}^\dagger(\vk ')] =-i(\gamma_R+pi\gamma_I)
\frac{l_P^2}{2}\delta_{rs}\delta_{p{\bar q}} \delta(\vk-\vk ')\; ,
\ee
where ${\overline q}=-q$. Compared to~\cite{paper},
the factor $\gamma p$ has been replaced by $\gamma_R+pi\gamma_I$.
This is algebraic novelty number three, the last one in our calculation.
For real $\gamma$ the $p$ dependence is erased from the 
commutation relations.

\section{The Hamiltonian}
We now have all the ingredients to find a Hamiltonian in terms of graviton
creation and annihilation operators (which will be linear combinations of
the perturbations in the metric and connection variables). A surprise
is in store at this point: in spite of the three novelties in the ingredients,
spelled out above, the final result for the graviton operators and Hamiltonian
is formally the same.

The gravitational Hamiltonian in terms of Ashtekar variables is given by
\bea \label{HamAsh} {\cal H}&=&\frac{1}{2l_P^2}\int d^3x N E^a_i E^b_j
\Big[\epsilon_{ijk}(F^k_{ab}+H^2 \epsilon _{abc} E^c_k)\nn
&&-2(1+\gamma^2)K^i_{[a} K^j_{b]}\Big] \label{fullH}\eea where \be\label{extK}
K^i_a=\frac{A^i_a-\Gamma^i_a(E)}{\gamma} \ee is the extrinsic
curvature of the spatial surfaces. The total Hamiltonian 
includes two further constraints,
the Gauss and vector constraint, but they are automatically satisfied by
expansions (\ref{fourrier}) and do not contribute to the order in perturbation theory we will consider 
\cite{paper}. The dynamics of the theory is encoded by
the second order Hamiltonian quadratic in first order perturbations.
To derive this Hamiltonian, a number of subtleties need
to be taken into account which are 
spelled out in detail in \cite{paper}. To write the Hamiltonian as a product of graviton creation
and annihilation operators inside the horizon, we need to express the second order Hamiltonian in terms of the mode expansion (\ref{fourrier}) (see Appendix III of \cite{paper}). 

We can determine the graviton operators inside the horizon ($|k\eta| \gg 1$) 
following the same procedure as in~\cite{paper}. Before reality conditions 
are imposed there should be unphysical modes that vanish on-shell
(and that will turn out to have negative energy and norm). 
The physical modes should commute with the non-physical modes 
and reduce, on-shell, to the correct expressions in terms of metric variables. 
Using these rules, and recalling (\ref{modesonshell+}) 
and (\ref{fixedcrs}), we define
\be \label{Gr+}G_{r{\cal P_+}} =
\frac{-r}{i\gamma}{\left({\tilde a}_{r+}- k(r+i\gamma){\tilde
e}_{r+}\right)}\ee\be \label{Gr-}
G_{r{\cal P_-}}=\frac{-r}{i\gamma}({\tilde a}_{r+}- k(r-i\gamma){\tilde
e}_{r+})\ee
\be
\label{Gr+dagger}G^{\dagger}_{r{\cal P_+}}=\frac{r}{i\gamma}({\tilde
a}^\dagger_{r-}- k(r-i\gamma){\tilde e}^\dagger_{r-})\ee
\be \label{Gr-dagger}G^{\dagger}_{r{\cal P_-}}=\frac{r}{i\gamma}({\tilde
a}^\dagger_{r-}- k(r+i\gamma){\tilde e}^\dagger_{r-})\; . \ee
The index ${\cal P}={\cal P_+},{\cal P_-}$ denotes physical and unphysical
modes, respectively. The normalisation ensures the right behaviour on-shell,
i.e $G_{r{\cal P_-}}\approx 0$ and $G_{r{\cal P_+}}\approx 2rk e_r$. 
Once the reality conditions (\ref{totreal})--(\ref{totrealdagger})--(\ref{realg}) are enforced  
one can check that the $G^\dagger$ are indeed
the Hermitian conjugate operators of the $G$. The commutation relations
are, as required:
\bea\label{Galgebra}
\left[G_{r{\cal P}}(\vk),G^{\dagger}_{s{\cal P}}(\vk')\right]&=&{\cal P} k l_P^2
\delta_{rs}
\delta(\vk-\vk')\\
\left[G_{r{\cal P_+}}(\vk),G^{\dagger}_{s{\cal P_-}}(\vk')\right]&=&0\; . \eea
These expressions are precisely the same as found in~\cite{paper} for
a purely imaginary $\gamma$, in spite of the three algebraic novelties spelled
out above. Somehow the modifications conspire to give the same graviton 
operators and commutators between them. This means that the Hamiltonian 
in terms of graviton states can be written in the same way as equation 
(105) of~\cite{paper}. Just like before an inner product, enforcing the
reality conditions, may be found in the representation diagonalizing
the $G^\dagger$ operators. The state ${\cal P}={\cal P_+}=1$ has positive energy and norm,
and ${\cal P}={\cal P_-}=-1$ has negative energy and norm. 
On-shell, the Hamiltonian becomes:
\be \label{Hamshell}{\cal H}^{ph}_{eff}\approx
\frac{1}{2l_P^2}\int d\vk \sum_r   [{G}^{ph}_{r}{ G}^{ph
\dagger}_{r} (1+ir\gamma)+ {G}_{r}^{ph \dagger}{G}^{ph}_{r}
(1-ir\gamma)]\\ \ee where ${G}^{ph}_{r}=G_{r {\cal P_+}}$.

The first term in the Hamiltonian we have just derived (which 
follows from a EEF ordering) 
needs to be normal ordered, leading to a
chiral (i.e. $r$-dependent) vacuum energy $V_r\propto 1+ir\gamma$.  
The chiral asymmetry is given by
\be\label{chiraleq}
\frac{V_R-V_L}{V_R+V_L}=i\gamma\; . \ee
In~\cite{paper} it was found that for imaginary $\gamma$ the 
vacuum energy (VE) is chiral and that for $|\gamma|>1$ one of the modes 
has negative VE. This flags a point of interest, since a 
negative VE is usually associated with fermionic degrees
of freedom. We now find that for $\gamma$ with a real part the VE
for each mode is complex. The imaginary part, however, is maximally chiral
and so cancels out, when right and left modes are added together. The real part
never sees such a cancellation, except in the limit when $|\Im(\gamma)|
\rightarrow\infty$, and so the total VE is only zero for the Palatini-Kibble
theory.

What is the origin of this result? We already pointed out in~\cite{paper}
that non-perturbatively the Hamiltonian is generally complex,
a matter behind many of the novelties we have exposed. On-shell
the Hamiltonian is zero and therefore real. The complexity of the 
Hamiltonian is not to be confused with its Hermiticity
after quantization and the inner product should enforce the
Hermiticity of the quantum
Hamiltonian. Perturbatively, however, the situation 
is more complicated. As explained in~\cite{paper}, even though the 
second order Hamiltonian must still
be zero on-shell, the portion dependent on first order variables 
(to be seen as the perturbative Hamiltonian ${\cal H}^{eff}$) evades
the Hamiltonian constraint. A number of other novelties of this sort 
appear when going from the full theory to perturbation theory. It turns
out that the classical perturbed Hamiltonian is always real on-shell,
even if it's no longer zero. This is still true for a generally complex
$\gamma$. However, quantum mechanically the perturbative Hamiltonian is 
only Hermitian, on and off-shell, 
{\it if $\gamma$ is imaginary}. If $\gamma$ has a real part the 
normal ordered Hamiltonian is still Hermitian, but the VE is not.
This can easily be seen from (\ref{Hamshell}): obviously 
$G_r^{ph}G_r^{ph\dagger}$ and $G_r^{ph\dagger}G_r^{ph}$ are still Hermitian
under the chosen inner product, but their coefficients spoil Hermiticity
before, but not after ordering. 

What attitude should we take towards this result? One possibility is that 
there's nothing wrong with it. Obviously the VE couples to the Einstein's 
equations, but the total is always real. Should we decide, however, that this
feature is pathological then there are two possible implications. 
One is that a purely imaginary $\gamma$ should be favoured. Another is 
that a symmetric ordering of the Hamiltonian constraint is to be preferred.
For more detail on the different ordering prescriptions see
\cite{paper}; however it's obvious that  $EFE$ or 
$\frac{1}{2}\left(EEF+FEE\right)$ ordering would satisfy 
${\cal H}={\cal H^\dagger}$ on and off-shell, before and after ordering.
In this case there would be no chirality in the VE;
however, as the graviton modes are still the same, the vacuum fluctuations, or the 2-point function,
would still exhibit a chiral signature, as
investigated in the next section.

\section{Vacuum fluctuations}\label{fluct}

As in \cite{paper}, we now want to compute the 2-point function in terms of connection variables as it
determines the vacuum fluctuation power spectrum. This is given by 
\be \label{PS1}{\langle
0|A^\dagger_r(\vk)A_r(\vk')|0\rangle} =P_r(k)\delta(\vk-\vk')\;
,\ee where $A_r(\vk)$ represents Fourier space connection variables with handedness $r$, i.e.
\be\label{Bigak} A_r(\vk)=a_{r+}(\vk) e^{-i k\cdot x} +
a_{r-}^\dagger(\vk) e^{i k\cdot x}\; . \ee Note that (\ref{PS1}) depends on a specific ordering of the 2-point 
function, and in general we have to consider \be A^\dagger A \rightarrow
\alpha A^\dagger A + \beta A A^\dagger\; , \ee with
$\alpha+\beta=1$ and $\alpha,\beta>0$. As (\ref{PS1}) is a variance,
it must always be real and positive (as opposed to the vacuum energy). Any chiral effects will then
leave a measurable imprint on this 
quantity.

\begin{figure}
\begin{center}
\includegraphics[width=6cm]{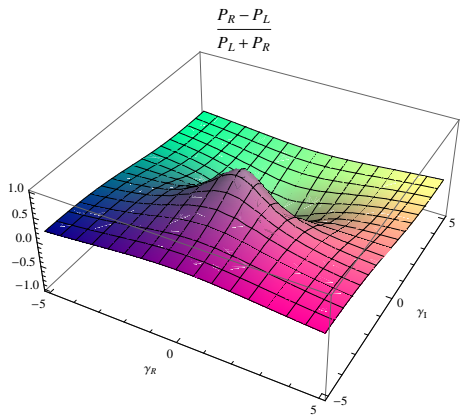}
\end{center}
\caption{Power spectrum asymmetry as a function of a generally
complex Immirzi parameter $\gamma$.}
\label{chirality}
\end{figure}

We need to relate the power spectrum to the physical graviton modes found in section IV.
This can be done by substituting the on-shell conditions (\ref{modesonshell-}) into 
(\ref{Gr+}) and (\ref{Gr+dagger}):
\bea
a^{ph}_{r+}&=&\frac{r-i\gamma}{2r}G_{r{\cal P_+}}\\
a^{ph\dagger }_{r+}&=&\frac{r+i\gamma^\ast}{2r}G^\dagger_{r{\cal P_+}}\\
a^{ph}_{r-}&=&\frac{r-i\gamma^\ast}{2r}G_{r{\cal P_+}}\\
a^{ph\dagger}_{r-}&=&\frac{r+i\gamma}{2r}G^\dagger_{r{\cal P_+}}\;
. \eea Plugging these expressions into (\ref{Bigak}) 
we obtain: \bea A^{ph}_r(\vk)&=&\frac{r-i\gamma}{2r}G_{r{\cal
P_+}}(\vk) e^{-i k\cdot x} +
\frac{r+i\gamma}{2r} G_{r{\cal P_+}}^\dagger(\vk) e^{i k\cdot x} \nonumber \\
A_r^{ph\dagger}(\vk)&=&\frac{r-i\gamma^\ast}{2r}G_{r{\cal P_+}}(\vk)
e^{-i k\cdot x} + \frac{r+i\gamma^\ast}{2r} G_{r{\cal
P_+}}^\dagger(\vk) e^{i k\cdot x}\nonumber \eea 
so that 
\be\label{2point}
{\langle
0|A^{ph\dagger}_r(\vk)A^{ph}_r(\vk')|0\rangle}=P_r(\gamma)
{\langle 0|G_{r{\cal P_+}}(\vk)G^\dagger_{r{\cal
P_+}}(\vk')|0\rangle}\; , \ee
where
\be
P_r(\gamma)=\frac{(r+i\gamma)(r-i\gamma^\ast)}{4}
=\frac{1-2\gamma_I r+|\gamma|^2}{4}\; . \ee
If $\gamma_Ir<0$, $P_r(\gamma)$ is obviously
positive. Otherwise, 
\be P_r(\gamma)\propto 1-2|\gamma_I|
+\gamma_I^2+\gamma_R^2=(1-|\gamma_I|)^2+\gamma_R^2 \ee
so this is also positive for any complex $\gamma$. Therefore the
2-point function is indeed always real and positive, as required.
The chiral asymmetry in the power spectrum can be expressed as
\be\label{chiralP}
\frac{P_R-P_L}{P_R+P_L}=-\frac{2\gamma_I}{1+|\gamma|^2} \;,\ee
or, for a general ordering, \be\frac{P_R-P_L}{P_R+P_L}=\frac{2(\beta-
\alpha)\gamma_I}{1+|\gamma|^2}\;. \ee 
This implies that for a real $\gamma$ there is no asymmetry
in the vacuum fluctuations for right and left gravitons. The chirality clearly traces to the fact that for
an imaginary $\gamma$ there must exist graviton and anti-graviton modes, i.e. the connection
is a complex field. Note however that the presence of a real part of the Immirzi parameter does
affect the {\it absolute} value of the asymmetry due to the factor $|\gamma|$ in the denominator of (\ref{chiralP}).
The power spectrum asymmetry (\ref{chiralP}) is plotted against a range of
values of $\gamma$ figure (\ref{chirality}). It is obviously antisymmetric in $\gamma_I$, the minimum and maximum 
being at $\gamma=\pm i$ respectively which are the values that correspond to a SD/ASD connection. They display the maximum chirality because the Palatini action can naturally be split into a SD and ASD part \cite{thbook}. The axis $\gamma_I=0$ corresponds to a real $\gamma$ and therefore no asymmetry.
The chirality also vanishes in the limit
$|\gamma|\rightarrow \infty$ which corresponds to the
Palatini-Kibble theory.

\section{A purely  real $\gamma$}\label{realgamma}
In everything we have derived so far we can take the limit $\Im(\gamma)
\rightarrow 0$ and regard the result as the real theory. The question
remains as to whether this limit is the same as a purely real theory,
in which all the variables are real from the start. In principle the 
two might be different, since some aspects of the construction are 
obviously discontinuous. For example, in a purely real theory expansions
(\ref{fourrier}) have modes $a_r$ and $e_r$ without a $p$ index, 
so that for a fixed $\vk$ and $r$ we start off with two, rather 
than four modes. It is important to check that this discontinuity 
does not propagate into our results, leading to expressions 
different from those taking the limit  
$\Im(\gamma) \rightarrow 0$ in the complex theory. In this Section
we show that this is not the case: at the very least it
is possible to set up the real theory so that no discontinuities
arise in any of the expressions in this paper, even though there is a jump in the number of 
independent degrees of freedom. Note that this is far 
from obvious since the statements 
$ {\tilde e}_{r+}={\tilde e}_{r-}$ and 
${\tilde a}_{r+}={\tilde a}_{r-}$
are second class constraints in the complex theory, and are not enforced
as operator conditions, but as formal conditions on the inner product.
The real theory results from imposing them as operator conditions. 

Firstly, the commutation 
relations (\ref{fixedcrs}) continuously shrink to
\be [{\tilde a}_{r}(\vk),{\tilde
e}_{s}^\dagger(\vk ')] =-i\gamma
\frac{l_P^2}{2}\delta_{rs} \delta(\vk-\vk ')\; 
\ee
for a real $\gamma$.
The reality conditions (\ref{totreal})--(\ref{totrealdagger})--(\ref{realg})
are now trivial (stating $0=0$) and do not constrain the theory.
However, for graviton operators we can still use definitions
(\ref{Gr+})--(\ref{Gr-dagger}) for $G_{r{\cal P}}$, simply dropping the 
$p$ index from their right-hand side, for example:
\be \label{Gr+1}G_{r{\cal P_+}} =
\frac{-r}{i\gamma}{\left({\tilde a}_{r}- k(r+i\gamma){\tilde
e}_{r}\right)}\; .\ee
It may appear that we are
introducing too many modes. In the complex theory, for a fixed $\vk$ and
$r$ we start with four modes, ${\tilde a}_{rp}$ and ${\tilde e}_{rp}$,
from which we build four $G_{r{\cal P}}$ and $G_{r{\cal P}}^\dagger$.
Three reality and torsion-free
conditions then reduce them to a single physical operator,
as explained after Eqn.~(\ref{totrealdagger}). For the real theory
we only have two modes, $a_r$ and $e_r$, from which we build four
$G_{r{\cal P}}$ and $G_{r{\cal P}}^\dagger$ without having any reality conditions.
However, upon closer inspection we see that for fixed
$\vk$ and $r$ there are only two independent
modes among the $G_{r{\cal P}}$ and $G_{r{\cal P}}^\dagger$: In the complex
theory we needed the reality conditions to ensure that the
$G_{r{\cal P}}^\dagger$ were in fact the Hermitian conjugates of the 
$G_{r{\cal P}}$. If we drop the index $p$ from their expressions, as in 
(\ref{Gr+1}), then this fact follows trivially from their definitions
and the linearity of the $\dagger$ operation. Hence by defining 
gravitons operators in the real theory we do preserve the number of
independent degrees of freedom.

The issue persists on how to eliminate the non-physical mode. This is 
done by imposing the torsion-free condition, relating the $a_r$ to the 
$e_r$, which amounts to disqualifying the $G_{r{\cal P_-}}$ mode. A
{\it possible} implementation, even in the real theory, is to do this via
the inner product. 
As in~\cite{paper}, we work in a holomorphic representation which
diagonalizes $G^\dagger_{r{\cal P}}$, i.e.: $ G^\dagger_{r{\cal P}}\Phi(z)=
z_{r{\cal P}}\Phi(z)$. Then  (\ref{Galgebra}) implies: 
\be \label{Grop}
G_{r{\cal P}}\Phi = {\cal P}k l_P^2 \frac{\partial \Phi}{\partial z_{r{\cal P}}} \;
.\ee
Following the same procedure as in~\cite{paper} we find
\be\label{ansatz}{\langle
\Phi_1 | \Phi_2\rangle}=\int d z d {\bar {z}} e^{\mu(z,{\bar z})}
{\bar \Phi_1}({\bar z}) \Phi_2 (z)\ee with:
\be \mu(z,{\bar
z})=\int d{\vk}\sum_{r{\cal P}}\frac{-{\cal P}}{k l_P^2}z_{r{\cal P}}(\vk){\bar z}
_{r{\cal P}}(\vk)\; , \ee 
rendering the states built from operators with 
${\cal P}={\cal P_-}=-1$ non-normalizable. 
As long as this procedure is adopted for the real theory the expressions
found in this paper are continuous, and the limit 
$\Im(\gamma)\rightarrow 0$ does indeed represent the real theory.

\section{Conclusion}

In this paper we have generalized the results of \cite{paper} to cover all values of the
Immirzi parameter. Our analysis shows that an imaginary part of $\gamma$ is needed to produce
a chiral effect in the vacuum fluctuations, whereas a purely real $\gamma$ would give 
the same physical Hamiltonian for right- and left-handed gravitons. The greatest asymmetry
occurs for the values $\gamma=\pm i$, corresponding to a SD/ASD connection and
the subject of \cite{prl}. Here, as in previous work, the chirality also depends on the ordering used for the 2-point function. Although this implies
that an observation of this asymmetry cannot be traced back to one single cause, it is still
a striking prediction of quantum gravity in the Ashtekar formalism.

It was shown in~\cite{TBpapers}
that even a small chiral effect in the gravitational wave background would greatly simplify its
detection, making us hopeful that a test of our prediction could even be achieved by PLANCK. 
Note that other mechanisms exist that produce 
a similar chiral effect~\cite{steph,gianl,merc}, but the one pointed out here is by far the simplest.
It would be interesting to make contact with the work of~\cite{recentgravitons}, where a chiral
contribution was found for the graviton propagator. However, in this publication a Euclidean
signature and a real $\gamma$ were used, basically the opposite of our set-up, making the link
between the predictions unclear.

{\bf Acknowledgements}
We thank Dionigi Benincasa, Gianluca Calcagni and Chris Isham for 
help regarding this project.



\begin{thebibliography}{99}
\bibitem{pulbook}R. Gambini and J. Pullin , Loops, Knots, Gauge theories
and Quantum gravity, CUP, Cambridge 1996.
\bibitem{rovbook}C. Rovelli, Quantum Gravity, CUP, Cambridge, 2004.
\bibitem{thbook}T. Thiemann, Modern Canonical Quantum General Relativity,
CUP, Cambridge, 2007.
\bibitem{recentgravitons}
C. Rovelli, Phys. Rev. Lett. 97, 151301, 2006; E. Bianchi et
al, Class. Quant. Grav. 23, 6989, 2006; E. Bianchi, E. Magliaro,
C.Perini, Nuc. Phys. B 822, 245, 2009.
\bibitem{rovrecent}C. Rovelli, arXiv:1004.1780 and 1010.1939.
\bibitem{prl}J. Magueijo and D. Benincasa,
Phys. Rev. Lett. 106: 121302, 2011.
\bibitem{paper} L. Bethke and J. Magueijo,  arXiv:1104.1800.
\bibitem{muk}V. Mukhanov, ``Physical Foundations of cosmology'', CUP, Cambridge
2005.
\bibitem{lid}A. Liddle and D. Lyth, ``Cosmological Inflation and Large-scale
Structure'', CUP, Cambridge 2000.
\bibitem{gravitons}A. Ashtekar, C. Rovelli and L. Smolin, Phys. Rev. D44,
1740, 1991.
\bibitem{leelaur}L. Freidel and L. Smolin, Class. Quant. Grav. 21: 3831-3844,
2004.
\bibitem{TBpapers}C. Contaldi, J. Magueijo and L. Smolin,
Phys.Rev.Lett. 101: 141101, 2008.
\bibitem{steph}S. Alexander, arXiv:0706.4481.
\bibitem{gianl}S. Alexander and G. Calcagni, Found.Phys.38, 1148-1184, 2008;
Physics Letters B 672 (2009) 386.
\bibitem{merc}S. Mercuri, arXiv:1007.3732.

\end{thebibliography}
\end{document}